\documentclass[12pt]{article}
\pdfoutput=1
\usepackage{graphicx,psfrag,epsf,color}
\usepackage[dvipsnames]{xcolor}
\usepackage{amsmath,amssymb,amsfonts, booktabs}
\usepackage{appendix}
\usepackage{multirow}
\usepackage{geometry}
\usepackage{float}
\usepackage{mathrsfs}  
\usepackage{authblk}
\usepackage{array}
\usepackage{cite}
\usepackage{slashed}
\usepackage{graphicx}
\usepackage{adjustbox}
\usepackage{dsfont}
\graphicspath{ {Img/} }
\usepackage{rotating}
\usepackage{soul}
\usepackage{slashed, cancel}
\usepackage{bbold}
\usepackage[caption=false]{subfig}
\usepackage{hyperref}
\usepackage{comment}
\usepackage{soul}
\newcounter{MBQ}

\usepackage{tabularx}
\newcolumntype{C}{>{\centering\arraybackslash}X}

\newcommand{\alem}{\alpha_{\rm em}}
\newcommand{\as}{\alpha_s}
\newcommand{\LamQCD}{\Lambda_{\rm QCD}}

\newcommand{\nub}{{\bar{\nu}_\ell}}

\newcommand{\red}[1]{{\color{blue}#1}}

\numberwithin{equation}{section}

\begin{document}
\allowdisplaybreaks

\begin{titlepage}

\begin{flushright}
{\small
January 22nd, 2026
}
\end{flushright}

\vskip1cm
\begin{center}
{\Large \bf\boldmath One-Loop QCD Corrections to $\bar{B}\to X_c \ell \bar{\nu}_\ell$\\ 
in and Beyond the Standard Model}
\end{center}

\vspace{0.5cm}
\begin{center}
{\sc Alexandre Carvunis,$^a$ \sc Gael Finauri$^b$} \\[6mm]
{\it{$^a$TU Dortmund University, Department of Physics, \\ Otto-Hahn-Str.4, D-44221 Dortmund, Germany} \\ \small \tt{\href{mailto:alexandre.carvunis@tu-dortmund.de}{alexandre.carvunis@tu-dortmund.de}}}\\[6mm]
{\it $^b$Dipartimento di Fisica, Università di Torino \& INFN, Sezione di Torino,\\
Via Pietro Giuria 1, I-10125 Turin, Italy \\ \small \tt{\href{mailto:gael.finauri@unito.it}{gael.finauri@unito.it}}}

\end{center}

\vspace{0.6cm}
\begin{abstract}
\vskip0.2cm\noindent

We compute one-loop QCD corrections to the triple differential width of the inclusive decay $\bar{B} \to X_c \ell \nub$ including contributions from all relevant dimension-six operators in the Weak Effective Theory (WET), at leading power in heavy quark expansion.
Furthermore we derive for the first time up to order $\mathcal{O}(\as)$ analytic expressions for the first three moments of the distribution in the lepton energy, hadronic invariant mass and dilepton invariant mass, in the presence of beyond the Standard Model contributions from the WET. 
\end{abstract}

\end{titlepage}

%%%%%%%%%%%%%%%%%%%%%%%%%%%%%%%%%%%%%%%%%%%%%%%%%%%%%%%%%%%%%%%%%%%
\tableofcontents
\newpage

\section{Introduction}
The available data on inclusive semileptonic $\bar{B}$ decays is opening up the possibility to study for the presence of New Physics (NP) in $b \to c \ell \nub$ transitions.
The natural framework for the phenomenology of $b$-hadrons in the Standard Model (SM) is the weak effective theory (WET), in which all the degrees of freedom heavier than the $b$ quark are integrated out. 
The same framework can be used for NP if we assume that the latter only exists at a scale far above the $b$ quark mass. In this case, the effects of any NP scenario below the $b$ quark scale can be fully described in the WET, which considers all possible operators of mass dimension-six (and higher) which respect the $\text{SU}(3)_c \times U(1)_{\rm em}$ gauge symmetry.
This is possible because the physics of energies above the $b$ quark mass is integrated out and stored into Wilson coefficients (WC) multiplying the WET operators through what is called a matching procedure.
Therefore, when extracting WC directly from experimental data, any significant deviation from the known Standard Model values of these WC would be a smoking gun for physics beyond the Standard Model (BSM). In addition, patterns observed in BSM WC could help us infer the nature of New Physics at high energy.

For these reasons, BSM effects in the inclusive decay $\bar{B} \to X_c \ell \nub$ have been actively studied~\cite{Czarnecki:1992zm,Grossman:1995yp,Dassinger:2007pj,Dassinger:2008as,Colangelo:2016ymy,Kamali:2018bdp,Colangelo:2020vhu,Fael:2022wfc}, and have been used to put bounds on specific NP models~\cite{Feger:2010qc,Crivellin:2014zpa,Celis:2016azn,Kamali:2018fhr,Jung:2018lfu,Iguro:2020cpg}. 
In our recent work \cite{Carvunis:2025vab} we presented the first global fit to $\bar{B} \to X_c \ell \nub$ observables with generic NP operators and found bounds on the BSM Wilson coefficients competitive with those coming from exclusive $b \to c \ell \nub$ modes. 
Until further progress is made in lattice QCD \cite{Gambino:2020crt,Gambino:2022dvu,Barone:2023tbl,DeSantis:2025qbb,DeSantis:2025yfm,Kellermann:2025pzt}, the most accurate way to obtain predictions for observables in inclusive semileptonic $B$ decays is through a double expansion of the forward scattering matrix of the $B$ meson in a local Operator Product Expansion (OPE)~\cite{Blok:1993va,Manohar:1993qn,Bigi:1993fe,Gremm:1996df}, yielding contributions going as $\mathcal{O}((\LamQCD/m_b)^i \times \alpha_s^j)$. 
In \cite{Carvunis:2025vab}, our calculation includes power corrections up to $\mathcal{O}(\LamQCD^3/m_b^3)$, and perturbative corrections up to $\mathcal{O}(\alpha_s^2)$ in the SM. In the BSM terms the $\mathcal{O}(\alpha_s)$ corrections were only included for the right-handed vector operator as it is the only one interfering with the Standard Model (SM) (and hence linear in the NP Wilson coefficient). %\cite{Carvunis:2025vab}.
In this paper we supplement the calculation of Ref.~\cite{Carvunis:2025vab} by extending the $\mathcal{O}(\as)$ corrections to the whole set of relevant dimension-6 WET operators.
Due to the scale hierarchy $m_b \sim m_c \gg \LamQCD$, the leading power of the OPE is given by the partonic decay $b \to c \ell \nub$, and it is hence independent on hadronic non-perturbative inputs.

The triple differential rate obtained through the OPE has to be interpreted as a distribution, valid only under the integration sign.
It is hence convenient to perform the comparison with experimental data only on integrated quantities.
We will therefore compute expressions for the measured observables, namely kinematic moments in: the lepton energy $E_\ell$, hadronic invariant mass $m_X^2$ (both measured with a lower cut in the lepton energy in the $\bar{B}$ rest frame)~\cite{BaBar:2004bij,CLEO:2004bqt,CDF:2005xlh,DELPHI:2005mot,Belle:2006jtu,Belle:2006kgy,BaBar:2009zpz} and dilepton invariant mass $q^2$ measured with a lower cut on $q^2$~\cite{Belle:2021idw,Belle-II:2022evt}.
The $q^2$ moments where also measured by CLEO a long time ago with a lower cut on the lepton energy~\cite{CLEO:2004bqt}, presenting puzzling tensions with the SM prediction~\cite{Finauri:2025ost}.
For this reason here we also consider such observables which could be measured in the near future by the Belle II experiment.

We present here for the first time analytic $\mathcal{O}(\as)$ results for the kinematic moments and the total width for the full set of NP operators in the WET.
To the best of our knowledge, in the SM analytic results at $\mathcal{O}(\as)$ are known only for the spectra~\cite{Jezabek:1988ja,Czarnecki:1989bz,Czarnecki:1994pu,Falk:1995me,Trott:2004xc,Aquila:2005hq} and partially for hadronic invariant mass moments~\cite{Falk:1997jq,Aquila:2005hq}.
Calculations of BSM effects at $\mathcal{O}(\as)$ have been performed in the past for specific operators, like the right-handed vector current~\cite{Dassinger:2007pj} and the scalar and pseudoscalar currents~\cite{Grossman:1995yp}(only for the total rate).
For the full set of NP operators, similar results to the ones presented in this work were obtained by numerical integration in Ref.~\cite{Fael:2022wfc}.
Their results, with which we only partially agree, were implemented in the open-source framework \texttt{Kolya}~\cite{Milutin:2024nbd}.

The paper is organized as follows:
we outline the theoretical framework of the OPE in Section~\ref{sec:theory}, while we describe the technical details of the one-loop calculation in Section~\ref{sec:calculation}.
In Section~\ref{sec:moms} we define the observables of interest in inclusive semileptonic $\bar{B}$ decays.
Finally we give some numerical results in Section~\ref{sec:num} before concluding in Section~\ref{sec:conclusion}.
The appendix contains numerical tables.

\section{Theoretical Framework}
\label{sec:theory}
We start by considering the weak effective Hamiltonian at dimension-six relevant for $b \to c \ell \nub$ transitions
\begin{equation}
\mathcal{H}_{\rm eff}= \frac{4 G_F}{\sqrt{2}} V_{cb} \sum_i C_i \mathcal{O}_i+\text {h.c.}\,,
\end{equation}
with $G_F = 1.1663787\cdot 10^{-5}~\text{GeV}^{-2}$~\cite{ParticleDataGroup:2024cfk} and $V_{cb}$ the Cabibbo-Kobayashi-Maskawa (CKM) matrix element factored out for convention.
We will consider for simplicity only left-handed neutrinos and lepton-flavour conservation. 
In cases where these assumptions do not hold one can still use the results presented in this paper with a few replacements~\cite{Carvunis:2025vab}.
The four-fermion operators $\mathcal{O}_i$ are given by
\begin{align}
\mathcal{O}_{V_L} &= [\bar{c}\, \gamma^\mu P_L b] \; [\bar{\ell}\, \gamma_\mu P_L \nu_{\ell}]\,, \qquad \mathcal{O}_{V_R} = [\bar{c}\, \gamma^\mu P_R b] \;[\bar{\ell}\, \gamma_\mu P_L \nu_{\ell}]\,, \nonumber\\
\mathcal{O}_{S} &= [\bar{c}\, b] \; [\bar{\ell}\, P_L \nu_{\ell}]\,, \qquad\qquad\quad\;\; \mathcal{O}_{P} = [\bar{c}\, \gamma_5 b] \; [\bar{\ell}\, P_L \nu_{\ell}]\,,\nonumber \\
\mathcal{O}_{T} &= [\bar{c}\, \sigma^{\mu \nu} P_L b] \; [\bar{\ell}\, \sigma_{\mu \nu} P_L \nu_{\ell}]\,,
\label{eq:NPopbasis}
\end{align}
with the chiral projectors $P_{L,R} = (1\mp \gamma^5)/2$ and the standard tensor Dirac structure $\sigma^{\mu\nu}=\frac{i}{2}[\gamma^\mu,\gamma^\nu]$.
Physics from scales larger than the $b$ quark mass $m_b$ is encoded in the dimensionless Wilson coefficients $C_i$.
In the SM all Wilson coefficients are zero except for $C_{V_L} = 1 + \mathcal{O}(\alem)$.
The inclusive decay is fully described by three kinematic variables.
They can be chosen to be the lepton and anti-neutrino energies ($E_\ell$ and $E_{\nub}$) in the $\bar{B}$ rest frame and the dilepton invariant mass $q^2 = (p_\ell + p_{\nub})^2$.
For massless leptons, the differential decay rate in these variables is~\cite{Carvunis:2025vab} 
\begin{align}
\label{eq:d3Gamma}
\frac{d^3 \Gamma}{d \hat{q}^2 d \hat{E}_{\ell} d \hat{E}_\nub} &= \frac{G_F^2|V_{cb}|^2 m_b^5}{\pi^3}\theta(\hat{E}_\ell) \theta(\hat{q}^2)\theta\Bigl(4\hat{E}_\nub \hat{E}_\ell - \hat{q}^2\Bigr)\biggl\{(w_0+2w_1)\frac{\hat{q}^2}{4}\nonumber\\
&+\frac{w_2}{4} (4 \hat{E}_\ell \hat{E}_\nub-\hat{q}^2)+\left(\frac{w_3}{2}- \text{Re}[w_{11}]\right) \hat{q}^2\left(\hat{E}_\nub-\hat{E}_\ell \right)\nonumber\\
& +w_{6} (8 \hat{E}_\ell \hat{E}_\nub-\hat{q}^2) +w_{7} \hat{q}^4 +2 w_{8}  \hat{q}^2 \left( \hat{E}_\ell+\hat{E}_\nub \right)
\biggr\}\,,
\end{align}
where the $w_i$ are scalar structure functions encoding the hadronic physics.
Here we introduced the notation of hatted variables which are normalized by appropriate powers of the $b$ quark mass, namely $\hat{q}^2 = q^2/m_b^2$, $\hat{E}_\ell = E_\ell/m_b$ and $\hat{E}_\nub = E_\nub/m_b$.
We will also use the normalized charm quark mass as $\rho = m_c^2/m_b^2$.
The structure functions are related to the discontinuity of the forward matrix element~\cite{Manohar:1993qn}
\begin{equation}
\label{eq:TGGp}
T_{\Gamma\Gamma'}(q) \equiv -\frac{i}{2m_B}\int d^4 x\, e^{-iq \cdot x} \langle \bar{B} | \text{T}\{J_{\Gamma'}^\dagger(x) J_\Gamma(0) \} |\bar{B}\rangle\,,
\end{equation}
where the currents are defined through
\begin{align}
    J_\Gamma(x) = \bar{c}(x) \Gamma M_\Gamma b(x)\,,
\end{align}
with the matrices $\Gamma = \{\mathds{1},\gamma^\mu, \sigma^{\mu\nu}\}$ and
\begin{equation}
  M_{\mathds{1}} = C_S + C_P \gamma^5\,, \qquad M_{\gamma^\mu} = C_{V_L}P_L + C_{V_R}P_R\,, \qquad M_{\sigma^{\mu\nu}} = C_T P_L\,.
\end{equation}
Depending on the number of Lorentz indices, one can tensor decompose~\eqref{eq:TGGp} as\footnote{See Ref.~\cite{Carvunis:2025vab} for more details about the tensor decomposition of the hadronic tensor.}
\begin{align}
\label{eq:tidec}
m_b T_{\mathds{1},\mathds{1}} &= t_0\,,\nonumber\\
m_b T_{\gamma^\mu, \gamma^\nu} &= -g^{\mu\nu} t_1+v^{\mu}v^{\nu}t_2+i\epsilon^{\mu\nu\alpha\beta}v_\alpha \hat q_\beta
t_3+\hat q^\mu \hat q^\nu t_4+(\hat q^\mu v^{\nu}+\hat q^\nu v^{\mu})t_5\,,\nonumber\\
m_b T_{\sigma^{\mu\nu}, \sigma^{\alpha\beta}} &= {\Sigma^{\mu\nu}}_{\kappa\lambda} {\Sigma^{*\alpha\beta}}_{\rho\tau}\biggl\{\left[v^\kappa\left(v^\tau g^{\lambda\rho}-v^\rho g^{\lambda\tau}\right)-v^\lambda\left(v^\tau
g^{\kappa\rho}-v^\rho g^{\kappa\tau}\right)\right]t_6 \nonumber \\ 
&\phantom{=}+\left[\hat q^\kappa\left(\hat q^\tau g^{\lambda\rho}-\hat
q^\rho g^{\lambda\tau}\right)-\hat q^\lambda\left(\hat q^\tau g^{\kappa\rho}-\hat q^\rho g^{\kappa\tau}\right)\right]t_7\nonumber\\
&\phantom{=}+\Big[g^{\kappa\tau}\left(v^\lambda \hat q^\rho+v^\rho\hat
q^\lambda\right)-g^{\lambda\tau}\left(v^\kappa \hat q^\rho+v^\rho\hat q^\kappa\right) \nonumber \\
& \phantom{=+} -g^{\kappa\rho}\left(v^\lambda \hat q^\tau+v^\tau\hat q^\lambda\right)+g^{\lambda\rho}\left(v^\kappa \hat q^\tau+v^\tau\hat q^\kappa\right)\Big]t_{8}\biggr\}\,,\nonumber\\
m_b T_{\mathds{1}, \gamma^\mu} &= \hat q^\mu t_9 + v^{\mu} t_{10}\,,\nonumber\\
m_b T_{\mathds{1}, \sigma^{\rho \tau}} &= \frac{i}{2}(v^{\rho}\hat{q}^\tau-v^{\tau}\hat{q}^\rho + i\epsilon^{\rho\tau\alpha\beta}v_\alpha \hat{q}_\beta)t_{11}\,,\nonumber\\
m_b T_{\gamma^\mu, \sigma^{\rho \tau}} &=
\frac{i}{2}(g^{\rho\mu}v^{\tau}-g^{\tau\mu}v^{\rho} +i \epsilon^{\mu\rho\tau\alpha}v_\alpha)t_{12}+\frac{i}{2}(g^{\rho\mu}\hat{q}^{\tau}-g^{\tau\mu}\hat{q}^{\rho} +i \epsilon^{\mu\rho\tau\alpha}\hat{q}_\alpha)t_{13}\nonumber\\
&+\frac{i}{2}(v^{\rho}\hat{q}^\tau-v^{\tau}\hat{q}^\rho + i \epsilon^{\rho\tau\alpha\beta}v_\alpha \hat{q}_\beta)(v^{\mu}t_{14}+\hat{q}^\mu t_{15})\,,
\end{align}
where $v = p_B/m_B$ stands for the $\bar{B}$ meson four-velocity, $\epsilon^{0123} = +1$ and we introduced the projector~\cite{Carvunis:2025vab}
\begin{equation}
    \Sigma^{\mu\nu\alpha\beta} = \frac{1}{4}\left(g^{\mu\alpha}g^{\nu\beta} - g^{\mu\beta}g^{\nu\alpha} - i\epsilon^{\mu\nu\alpha\beta} \right)\,.
\end{equation}
The structure functions $w_i$ are related to the $t_i$ through~\cite{Manohar:1993qn, Manohar:2000dt}
\begin{equation}
    w_i = \frac{i}{2\pi} \text{disc}[t_i]\,,
\end{equation}
and after a suitable change of variables they depend on only two of the three kinematic variables, $\hat{q}^2$ and the charm-quark off-shellness 
\begin{equation}
    \hat{u} = \frac{(m_b v-q)^2-m_c^2}{m_b^2}\,,
\end{equation}
which is related to the anti-neutrino energy as $\hat{E}_\nub = (1-\rho +\hat{q}^2-2\hat{E}_\ell -\hat{u})/2$.

Our goal is to compute the $w_i$ at order $\mathcal{O}(\as)$ at leading power in a $\LamQCD/m_b$ expansion.
At leading power in $\LamQCD/m_b$ the forward matrix element~\eqref{eq:TGGp} is given by the first term of a local operator product expansion (OPE), which can be computed partonically with on-shell $b$ quarks.

\section{Calculation}
\label{sec:calculation}
\begin{figure}
    \centering
    \includegraphics[width=0.3\textwidth]{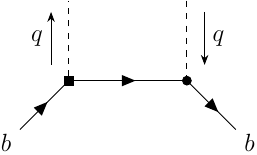}
    \caption{\small Tree level contribution to $T_{\Gamma\Gamma'}$ where the square dot denotes an insertion of $\Gamma M_\Gamma$ and the round dot denotes an insertion of $\gamma^0 M_{\Gamma'}^\dagger \Gamma^{\prime\dagger} \gamma^0$.}
    \label{fig:diagtree}
\end{figure}

\begin{figure}
    \centering
    \includegraphics[width=0.7\textwidth]{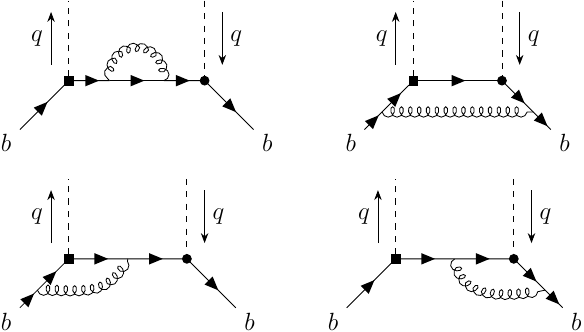}
    \caption{\small One-loop diagrams in QCD for $T_{\Gamma\Gamma'}$ where the square dot denotes an insertion of $\Gamma M_\Gamma$ and the round dot denotes an insertion of $\gamma^0 M_{\Gamma'}^\dagger \Gamma^{\prime\dagger} \gamma^0$. The dashed lines represent the injected momentum.}
    \label{fig:diagloop}
\end{figure}

The first step of the calculation is to compute the right-hand side of~\eqref{eq:TGGp} with external on-shell $b$ quark states of momentum $p_b = m_b v$ instead of the hadronic $\bar{B}$ meson states. 
The result for $T_{\Gamma \Gamma'}$ is then obtained with the simple replacement
\begin{equation}
\label{eq:proj}
    \bar{u}(p_b) M u(p_b) \to m_B\text{tr}\biggl[\frac{1+\slashed{v}}{2} M \biggr]\,,
\end{equation}
where $u(p_b)$ is the spinor for the external $b$ and $M$ a generic matrix in Dirac space.
The substitution~\eqref{eq:proj} ensures the correct QCD matrix elements
\begin{align}
    \langle \bar{B}| \bar{b} \gamma^\mu b |\bar{B}\rangle &= 2 m_B v^\mu\,,\nonumber\\
    \langle \bar{B}| \bar{b} \gamma^\mu \gamma^5 b |\bar{B}\rangle &= 0\,.
\end{align}
The matrix element between $b$ quark states is given at tree level by the single Feynman diagram shown in Figure~\ref{fig:diagtree}.
At one loop in QCD there are four diagrams (Figure~\ref{fig:diagloop}) to which we need to add the diagrams with counter-terms insertions.
In the cases where $T_{\Gamma \Gamma'}$ is a Lorentz tensor, before performing the loop calculation it is useful to project onto the scalar functions $t_i$.
This is done by inverting~\eqref{eq:tidec} and finding the Lorentz projectors $\mathcal{P}_i$ such that, for instance in the vector-vector case
\begin{equation}
    t_i = m_b \mathcal{P}^{\mu\nu}_i T_{\gamma^\mu \gamma^\nu}\,,
\end{equation}
and similarly for the others Dirac structures.
We provide the explicit form of the $\mathcal{P}_i$ in a \texttt{Mathematica} notebook contained in the ancillary files.

We employ dimensional regularization for both infrared (IR) and ultraviolet (UV) divergences with space-time dimensions $d = 4-2\epsilon$.
The effective operators~\eqref{eq:NPopbasis} are renormalized in the $\overline{\text{MS}}$ scheme~\cite{Aebischer:2017gaw}
\begin{equation}
   Z_{V_L,V_R} = 1\,, \qquad Z_{S,P} = 1 - \frac{\as C_F}{4\pi}\frac{3}{\epsilon} \,, \qquad Z_T = 1 + \frac{\as C_F}{4\pi} \frac{1}{\epsilon}\,,
\end{equation}
while the quark masses are in the on-shell scheme with the renormalization constant for the charm mass being
\begin{equation}
    Z_{m_c}^{\rm OS} = 1 - \frac{\as C_F}{4\pi} \biggl(\frac{3}{\epsilon} + 3\log \frac{\mu^2}{m_c^2} + 4\biggr)\,.
\end{equation}
The treatment of $\gamma^5$ is problematic in dimensional regularization as it is a strictly four-dimensional object.
We choose to work with the Breitenlohner-Maison~\cite{Breitenlohner:1977hr} implementation of the t'Hooft-Veltman scheme~\cite{tHooft:1972tcz} (BMHV) through the \texttt{Mathematica} package \texttt{FeynCalc}~\cite{Shtabovenko:2020gxv}.
In this scheme $\gamma^5$ is treated in four dimensions, as well as the external vectors and the Levi-Civita tensor.
The other Dirac matrices, the loop momenta and the metric tensor are separated into four-dimensional components and $(d-4)$ components.
This prescription artificially violates Ward identities, which have to be restored by adding finite counter-terms to operators containing $\gamma^5$
\begin{align}
\bar{c}\, \gamma^5 b \to Z_{5p} \bar{c}\, \gamma^5 b\,,\nonumber\\
\bar{c}\, \gamma^\mu\gamma^5 b \to Z_{5a} \bar{c}\, \gamma^\mu\gamma^5 b\,,\nonumber\\
\bar{c}\, \sigma^{\mu\nu}\gamma^5 b \to Z_{5t} \bar{c}\, \sigma^{\mu\nu}\gamma^5 b\,,
\end{align}
where at order $\mathcal{O}(\as)$ we find
\begin{equation}
    Z_{5p} = 1 -\frac{2\as C_F}{\pi}\,, \qquad Z_{5a} = 1 - \frac{\as C_F}{\pi}\,, \qquad Z_{5t} = 1\,.
\end{equation}
Note that the projectors $\mathcal{P}_i$ are purely four-dimensional objects.

After the contraction of the Lorentz indices with the projectors $\mathcal{P}_i$ and after the Dirac algebra has been simplified, the scalar integrals in the loop momentum $k$ are of the form
\begin{equation}
    I_{(a,b,c)} = \int \frac{d^d k}{(2\pi)^d} \frac{i(4\pi)^{2-\epsilon}e^{\epsilon \gamma_E} \mu^{2\epsilon} (m_b^2)^{a+b+c-2}}{(k^2+i\eta)^a (k^2 + 2p_b \cdot k +i \eta)^b (k^2 + 2(p_b -q) \cdot k +m_b^2 \hat{u}+ i\eta)^c}\,,
\end{equation}
where $a$, $b$ and $c$ can take integer values between -2 and 2.
Following the strategy employed in Ref.~\cite{Alberti:2012dn}, we perform the integration by parts (IBP) reduction with the help of \texttt{FIRE6}~\cite{Smirnov:2019qkx} and \texttt{LiteRed}~\cite{Lee:2008tj,Lee:2012cn,Lee:2013mka}, reducing this family of master integrals to a basis of five master integrals $I_{(0,0,1)}$, $I_{(0,1,0)}$, $I_{(0,1,1)}$, $I_{(1,0,1)}$ and $I_{(1,1,1)}$.
The explicit results for the master integrals are~\cite{Alberti:2012dn}
\begin{align}
\label{eq:masterseasy}
    I_{(0,0,1)} &= \left( \frac{\mu^2}{m_b^2}\right)^\epsilon e^{\epsilon \gamma_E} \Gamma(-1+\epsilon) \rho^{1-\epsilon}\,,\nonumber\\
    I_{(0,1,0)} &= \left( \frac{\mu^2}{m_b^2}\right)^\epsilon e^{\epsilon \gamma_E} \Gamma(-1+\epsilon) \,,\nonumber\\
    I_{(0,1,1)} &= -\frac{1}{\epsilon}-\log \frac{\mu^2}{m_b^2} -2 \\ 
    & + \frac{\sqrt{\rho}}{2(t-\sqrt{\rho})(t \sqrt{\rho}-1)}\biggl[t(t-2\sqrt{\rho})\log \rho + 2(t^2-1)\log t +\log \rho \biggr] + \mathcal{O}(\epsilon) \,, \nonumber
\end{align}
where the variable $t$ is defined such that
\begin{equation}
\label{eq:q2tot}
    \hat{q}^2 = \bigl(1-\sqrt{\rho}\, t\bigr)\biggl(1-\frac{\sqrt{\rho}}{t} \biggr)\,.
\end{equation}
The integrals in~\eqref{eq:masterseasy} are real, while $I_{(1,0,1)}$ and $I_{(1,1,1)}$ also carry an imaginary part.
The real and imaginary parts need to be treated differently as the real part will be multiplied by delta functions (and their derivatives) in $\hat{u}$ while the imaginary parts will be multiplied by inverse powers of $\hat{u}$.
Therefore in the imaginary parts it is essential to keep powers of $\hat{u}$ unexpanded in $\epsilon$, such that they will regulate endpoint divergences in the integrals over $\hat{u}$ (being then written in terms of plus distribution according to~\eqref{eq:plusdist}).
In the real parts on the other hand we can expand in $\epsilon$, but first it is convenient to expand the integrand around $\hat{u}=0$ before integration over Feynman parameters.
This is because otherwise there would be unregulated divergences coming from $\text{Re}[I_{(1,0,1)}]$ multiplied by $\delta^{(1)}(\hat{u})$.
After these considerations, for the real and imaginary part we find
\begin{align}
    \text{Re}[I_{(1,0,1)}] &= \biggl(\frac{\mu^2}{m_b^2}\biggr)^\epsilon\biggl[-\frac{2 \rho ^2+\hat{u}^2-\rho  \hat{u}}{2 \rho ^2 \epsilon} \nonumber \\ & \hspace{2.1cm} + \frac{-4 \rho ^2+\log \rho \left(2 \rho^2+\hat{u}^2-\rho  \hat{u}\right)-\hat{u}^2+2 \rho\hat{u}}{2 \rho ^2}\biggr] + \mathcal{O}(\hat{u}^3,\epsilon)\,,\nonumber\\
    \text{Im}[I_{(1,0,1)}] &=-\pi\theta(\hat{u})\left(\frac{\mu^2}{m_b^2}\right)^{\epsilon}e^{\epsilon \gamma_E} \frac{\Gamma(1+\epsilon)  \Gamma (1-\epsilon)^2}{\Gamma(2-2 \epsilon)}\hat{u}^{-\epsilon} \left(\frac{\hat{u}}{\rho+\hat{u}}\right)^{1-\epsilon}\,.
\end{align}

The integral $I_{(1,1,1)}$ is the more complicated and we will explain in more detail how to compute it.
After using Feynman parametrization one gets the following integral over the Feynman parameters
\begin{align}
    I_{(1,1,1)} &= \Gamma(1+\epsilon)e^{\epsilon \gamma_E} \biggl(\frac{\mu^2}{m_b^2}\biggr)^\epsilon \times \nonumber \\ & \hspace{1cm} \int_0^1 dx_1 \int_0^{1-x_1} dx_2 \biggl[x_1^2 +2x_1 x_2(1-\hat{q}_0) -\hat{u}x_2 + (\hat{u}+\rho)x_2^2-i\eta \biggr]^{-1-\epsilon}\nonumber\\
    &= \Gamma(1+\epsilon)e^{\epsilon \gamma_E} \biggl(\frac{\mu^2}{m_b^2}\biggr)^\epsilon \int_0^1 dx \int_0^1 dy\,  x^{-\epsilon} \biggl[x \chi(y,\hat{u}) -(1-y)(\hat{u} +i\eta) \biggr]^{-1-\epsilon}
\end{align}
where we performed the substitution $x_1 = x y$ and $x_2 = x (1-y)$ with $dx_1 dx_2 = x dx dy$ and defined the function
\begin{equation}
    \chi(y,\hat{u}) = \hat{q}^2 y^2 +(1-\rho -\hat{q}^2-\hat{u})y+\hat{u}+\rho\,.
\end{equation}
We split real and imaginary part of $I_{(1,1,1)}$ since the imaginary part will be multiplied by a $1/\hat{u}$ from the charm propagator, and therefore they need to be carefully handled by using the so-called plus distribution, defined in \eqref{eq:uplusdef}.
On the other hand the real part of $I_{(1,1,1)}$ will only contribute in terms containing a $\delta(\hat{u})$ and therefore we can safely set $\hat{u}=0$ to simplify the result.
\begin{align}
    \text{Re}[I_{(1,1,1)}] &= -\frac{1}{2}\biggl(\frac{\mu^2}{m_b^2}\biggr)^\epsilon \biggl(\frac{1}{\epsilon}I_1(0) - I_4(0) + \mathcal{O}(\epsilon) \biggr)\,,\nonumber\\
    \text{Im}[I_{(1,1,1)}] &= \biggl(\frac{\mu^2}{m_b^2}\biggr)^\epsilon \pi \hat{u}^{-2\epsilon} \theta(\hat{u}) \bigl(I_1(\hat{u}) + \epsilon (I_4(\hat{u})-2I_2(\hat{u})) + \mathcal{O}(\epsilon^2)\bigr)\,,
\end{align}
with
\begin{align}
    I_1(\hat{u}) &= \int_0^1 \frac{dy}{\chi(y,\hat{u})}\,, \qquad I_4(\hat{u}) = \int_0^1 dy \frac{\log \chi(y,\hat{u})}{\chi(y,\hat{u})}\,,\nonumber\\
    I_2(\hat{u}) &= \int_0^1 \frac{dy}{\chi(y,\hat{u})} \log(1-y)\,.
\end{align}
The results for the integrals are
\begin{align}
    I_1(\hat{u}) &= -\frac{\log \left(\frac{1-\hat{q}^2+\hat{u}+\rho -\sqrt{\hat{q}^4 - 2\hat{q}^2 (1+\rho+\hat{u})+(1-\rho-\hat{u})^2}}{1-\hat{q}^2+\hat{u}+\rho +\sqrt{\hat{q}^4 - 2\hat{q}^2(1+\rho + \hat{u})+(1-\rho
   -\hat{u})^2}}\right)}{\sqrt{\hat{q}^4-2 \hat{q}^2 (1+ \rho + \hat{u})+(1-\rho -\hat{u})^2}}\,,\nonumber\\
   I_2(0) &= \frac{t}{\sqrt{\rho} (1-t^2)}  \left[\text{Li}_2\left(1-\frac{\sqrt{\rho }}{t}\right)-\text{Li}_2\left(1-t \sqrt{\rho }\right)\right]\,,\nonumber\\
   I_4(0) &= \frac{2t}{\sqrt{\rho}(1-t^2)} \biggl[\text{Li}_2\left(\frac{t^2-1}{t \sqrt{\rho }-1}\right)-\text{Li}_2\left(\frac{\left(t^2-1\right) \sqrt{\rho }}{t \left(t \sqrt{\rho }-1\right)}\right) \nonumber\\ & \hspace{2.5cm} +\log
   \left(\frac{1-\sqrt{\rho}t}{1-\sqrt{\rho}/t}\right) \log \frac{\sqrt{\rho}}{t} -\log t\log \frac{\rho}{t}\biggr]\,,
\end{align}
where $I_2$ and $I_4$ will be only needed for $\hat{u}=0$.

The divergent terms for $\hat{u}\to 0$ are written in terms of plus distributions in order to extract the divergence in the form of a pole in $\epsilon$ through~\cite{Alberti:2012dn}
\begin{equation}
\label{eq:plusdist}
\hat{u}^{-A + B \epsilon} = \sum_{p=0}^{A-1} \frac{(-1)^p}{p!} \frac{\delta^{(p)}(\hat{u})}{1+p-A+B \epsilon} + \sum_{n=0}^\infty \frac{(B \epsilon)^n}{n!} \biggl[\frac{\log^n \hat{u}}{\hat{u}^A} \biggr]_+ \,,   
\end{equation}
for $A>0$ and $B \neq 0$ where the plus distribution is defined under integration as~\cite{Alberti:2012dn}
\begin{equation}\label{eq:uplusdef}
    \int d\hat{u} \biggl[ \frac{\log^n \hat{u}}{\hat{u}^A}\biggr]_+ f(\hat{u}) =  \int_0^1 d\hat{u} \frac{\log^n \hat{u}}{\hat{u}^A} \biggl( f(\hat{u}) - \sum_{p=0}^{A-1} \frac{\hat{u}^p}{p!} \frac{d^p f}{d\hat{u}^p}\biggl|_{\hat{u}=0} \biggr)\,.
\end{equation}

Finally, in order to get the structure functions $w_i$, we need to take the discontinuity of $t_i$ through the left-hand cut. 
This can be done by taking the imaginary part of the $t_i$, but considering the Wilson coefficients real.
For this purpose we use
\begin{equation}
    \text{Im}\biggl[\frac{1}{(\hat{u}+i\eta)^{n+1}} \biggr] = \pi \frac{(-1)^{n+1}}{n!}\delta^{(n)}(\hat{u})\,,
\end{equation}
where $\delta^{(n)}(\hat{u})$ is the $n$th derivative of the delta function. 
We provide the results for the $w_i$ at order $\mathcal{O}(\as)$ as ancillary \texttt{Mathematica} files, having checked that we fully reproduce the known SM results~\cite{Aquila:2005hq}.
Notice that we compute all of the 16 structure functions such that they can be used also for studying processes with a massive lepton.

\section{Observables from the Semileptonic Spectrum}
\label{sec:moms}
From the triple differential distribution~\eqref{eq:d3Gamma} we have to compute moments of the semileptonic decay rate. 
We will compute analytically, up to $\mathcal{O}(\as)$, moments in $q^2$, the lepton energy $E_\ell$ and the hadronic invariant mass $m_X^2$, with a lower cut ${E_\ell}_{\rm cut}$ on the lepton energy.
Furthermore we also compute $q^2$ moments with a lower cut $q^2_{\rm cut}$.
Our first step is to compute linear moments which serve as the building blocks for the normalized central moments.
The linear moments are explicitly defined as
\begin{align}
\label{eq:linearmom}
\mathcal{L}_n(\hat{E_\ell}_{\rm cut}) &= \frac{1}{\Gamma_0} \int_{\hat{E_\ell}_{\rm cut}}^{(1-\rho)/2}d \hat{E}_\ell (\hat{E}_\ell)^n \frac{d\Gamma}{d\hat{E}_\ell}\,,\nonumber\\
\mathcal{H}_n(\hat{E_\ell}_{\rm cut}) &= \frac{1}{\Gamma_0} \int_{\hat{E_\ell}_{\rm cut}}^{(1-\rho)/2} d\hat{E_\ell} \int d\hat{u}\, d\hat{q}^2 (\hat{m}_X^2)^n \frac{d^3 \Gamma}{d\hat{E}_\ell d\hat{u}\,d\hat{q}^2}\,,\nonumber\\
\mathcal{Q}_n(\hat{E_\ell}_{\rm cut}) &= \frac{1}{\Gamma_0} \int_{\hat{E_\ell}_{\rm cut}}^{(1-\rho)/2} d\hat{E}_\ell \int d\hat{u}d \hat{q}^2 (\hat{q}^2)^n \frac{d^3 \Gamma}{d\hat{E}_\ell d\hat{u} d\hat{q}^2}\,,\nonumber\\
\mathcal{Q}_n(\hat{q}^2_{\rm cut}) &= \frac{1}{\Gamma_0} \int_{\hat{q}^2_{\rm cut}}^{(1-\sqrt{\rho})^2}d \hat{q}^2 (\hat{q}^2)^n \frac{d\Gamma}{d\hat{q}^2}\,,
\end{align}
where we normalized by
\begin{equation}
\label{eq:Gamma0}
    \Gamma_0 = |V_{cb}|^2\frac{G_F^2 m_b^5}{192\pi^3}\,,
\end{equation}
for convenience.
The normalized hadronic invariant mass is given by
\begin{equation}
\hat{m}^2_X = \frac{1}{m_b^2}(p_B-q)^2 = \hat{\bar{\Lambda}}^2+(1+\rho)\hat{\bar{\Lambda}} +\rho +(1+\hat{\bar{\Lambda}}) \hat{u} - \hat{\bar{\Lambda}} \hat{q}^2\,,
\end{equation}
with the parameter
\begin{equation}
\hat{\bar{\Lambda}} = \frac{m_B - m_b}{m_b}\,,
\end{equation}
which we count as $\mathcal{O}(1)$ as done in previous analyses~\cite{Gambino:2013rza,Gambino:2011cq}.
From~\eqref{eq:linearmom} we also derive the observable $R^*$ and the total rate $\Gamma$ 
\begin{align}
R^*(\hat{E_\ell}_{\rm cut}) = \frac{\mathcal{L}_0(\hat{E_\ell}_{\rm cut})}{\mathcal{L}_0(0)}\,, \qquad\qquad \Gamma = \Gamma_0 \mathcal{L}_0(0)\,,
\end{align}

The lower $q^2$ cut in~\eqref{eq:linearmom} was suggested in~\cite{Fael:2018vsp} since it preserves reparametrization invariance (RPI).
Therefore, due to RPI, these observables depend on a restricted set of HQE parameters, which is particularly advantageous if one would go to higher orders in the power expansion~\cite{Mannel:2023yqf}.
On the other hand, $\mathcal{L}_n$ is intrinsically breaking RPI but provides some of the most constraining data in the phenomenological analysis.
Therefore, a combined measurement of the three different types of moments, all subject to a lower cut on the lepton energy, would be of interest in order to also have the experimental correlation between different moments.
For this reason we also provide the theoretical calculation for $q^2$ moments with a lower cut ${E_\ell}_{\rm cut}$.
As pointed out in~\cite{Finauri:2025ost}, the latter have only been measured by the CLEO collaboration more than 20 years ago.
These data presented puzzling tensions with their tree level theoretical predictions~\cite{Finauri:2025ost}, which we will update with $\mathcal{O}(\as)$ corrections in Section~\ref{sec:CLEO}.

\begin{figure}
    \centering
    \includegraphics[width=0.6\textwidth]{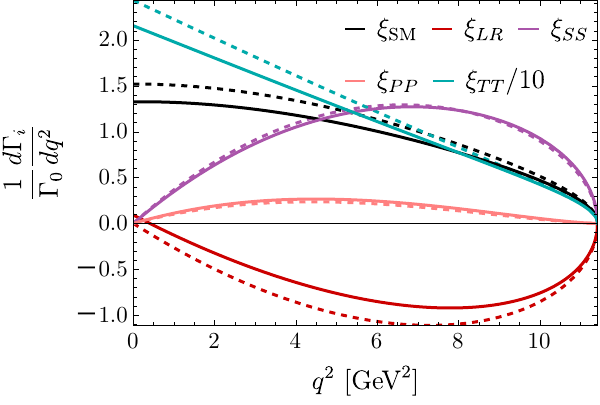}
    \caption{\small $q^2$ spectrum at LO (dashed) and NLO (solid). The contributions are labelled according to a decomposition analogous to~\eqref{eq:xidec}. The coefficient of the tensor term has been divided by 10 for presentation purposes, while the function $\xi_{RR}$ is not shown as it is identical to $\xi_{\rm SM}$.}
    \label{fig:q2spec}
\end{figure}

The calculation of $\mathcal{Q}_n(\hat{q}^2_{\rm cut})$ follows the same strategy employed in Section 3.4 of~\cite{Finauri:2025ost}.
The difference in the integral over $\hat{u}$ is that at leading power there are no derivatives of the delta functions. However at one-loop there are terms not proportional to delta functions which need to be integrated, making the result for the spectrum in $q^2$ more involved.
We show the result for the components of the spectrum in Figure~\ref{fig:q2spec}, where we adopted the inputs~\eqref{eq:inputs} of Section~\ref{sec:num}.

The spectrum is then integrated using the convenient change of variable~\eqref{eq:q2tot}~\cite{Fael:2024gyw} which maps the interval $\hat{q}^2 \in [0,(1-\sqrt{\rho})^2]$ onto $t \in [\sqrt{\rho},1]$.
We hence give our results for $\mathcal{Q}_n(\hat{q}^2)$ as functions of a lower cut on $t$
\begin{equation}
t_{\rm cut} = \frac{1+\rho-\hat{q}^2_{\rm cut}-\sqrt{(1+\rho-\hat{q}^2_{\rm cut})^2-4\rho}}{2\sqrt{\rho}}\,.
\end{equation}

For the rest of the observables we first perform the integral over $\hat{q}^2$, then the one over $\hat{u}$ and finally the remaining integral over $\hat{E}_\ell$ with a lower cut.
Analytic expressions for~\eqref{eq:linearmom} are available in \texttt{Mathematica} format from the authors upon request.

\subsection{Central Moments}
In this section we build normalized central moments, starting from the linear moments $\mathcal{M}_n =\{\mathcal{L}_n$, $\mathcal{H}_n,\mathcal{Q}_n\}$ computed in the previous section.
The normalized moments are easily obtained as $\hat{\mathcal{M}}_n \equiv \mathcal{M}_n/\mathcal{M}_0$, such that the observable is independent on the CKM input.
Already at this stage we re-expand the $\hat{\mathcal{M}}_n$ in $\as$ and the NP Wilson coefficients.
We define the first normalized moments by restoring the proper powers of $m_b$
\begin{equation}
\label{eq:firstmom}
    L_1 = m_b \hat{\mathcal{L}}_1\,, \qquad\qquad H_1 = m_b^2 \hat{\mathcal{H}}_1\,, \qquad\qquad Q_1 = m_b^2 \hat{\mathcal{Q}}_1\,.
\end{equation}
For the second and third moment it is customary to build central moments to reduce correlations among moments of different order.
They can be computed as the following linear combinations 
\begin{align}
\label{eq:centralmoms}
    L_n &= m_b^n\sum_{k=0}^n (-1)^{n-k} {n \choose k} \hat{\mathcal{L}}_k \hat{\mathcal{L}}_1^{n-k}\,,\qquad\qquad\;\; n\geq 2\,,\nonumber\\
    H_n &= m_b^{2n}\sum_{k=0}^n (-1)^{n-k} {n \choose k} \hat{\mathcal{H}}_k \hat{\mathcal{H}}_1^{n-k}\,,\qquad\qquad n\geq 2\,,\nonumber\\
    Q_n &= m_b^{2n}\sum_{k=0}^n (-1)^{n-k} {n \choose k} \hat{\mathcal{Q}}_k \hat{\mathcal{Q}}_1^{n-k}\,,\qquad\qquad n\geq 2\,.
\end{align}
The linear combinations are again re-expanded up to order $\mathcal{O}(\as)$ and to second order in the NP parameters.
Analytic results are provided in the ancillary files in the form of \texttt{Mathematica} expressions.

\subsection{Parametrization of New Physics Effects}
We reparametrize the complex Wilson coefficients as done in Ref.~\cite{Carvunis:2025vab}.
Since we are mostly interested in normalized observables, we define
\begin{equation}
    \tilde{C}_i = \frac{C_i}{C_{V_L}}\,,
\end{equation}
and use $a_{R,S,P,T}=|\tilde{C}_{V_R,S,P,T}|$ for the absolute values. For the three non-vanishing interference terms we define $\delta_R = \text{Arg}(\tilde{C}_{V_R})$, $\delta_{ST} = \text{Arg}(\tilde{C}_S\tilde{C}_T^*)$ and $\delta_{PT} = \text{Arg}(\tilde{C}_P\tilde{C}_T^*)$. This means that a generic NP scenario is described by 7 additional real parameters.
A normalized observable $\xi = \{R^*, L_n, H_n, Q_n\}$, after re-expanding up to $\mathcal{O}(a_i^2)$, will be of the form
\begin{align}
\label{eq:xidec}
    \xi &= \xi_{\rm SM} + a_R \cos(\delta_R) \xi_{LR} + a_R^2 \bigl(\xi_{RR} +\cos(\delta_R)^2 \xi_{LR^2}\bigr) + a_S^2 \xi_{SS} + a_P^2 \xi_{PP}\nonumber\\
    &\quad+ a_T^2 \xi_{TT} + a_S a_T \cos(\delta_{ST}) \xi_{ST} + a_P a_T \cos(\delta_{PT}) \xi_{PT}\,.
\end{align}

\section{Numerical Analysis}
\label{sec:num}
\begin{figure}
    \centering
    \includegraphics[width=\textwidth]{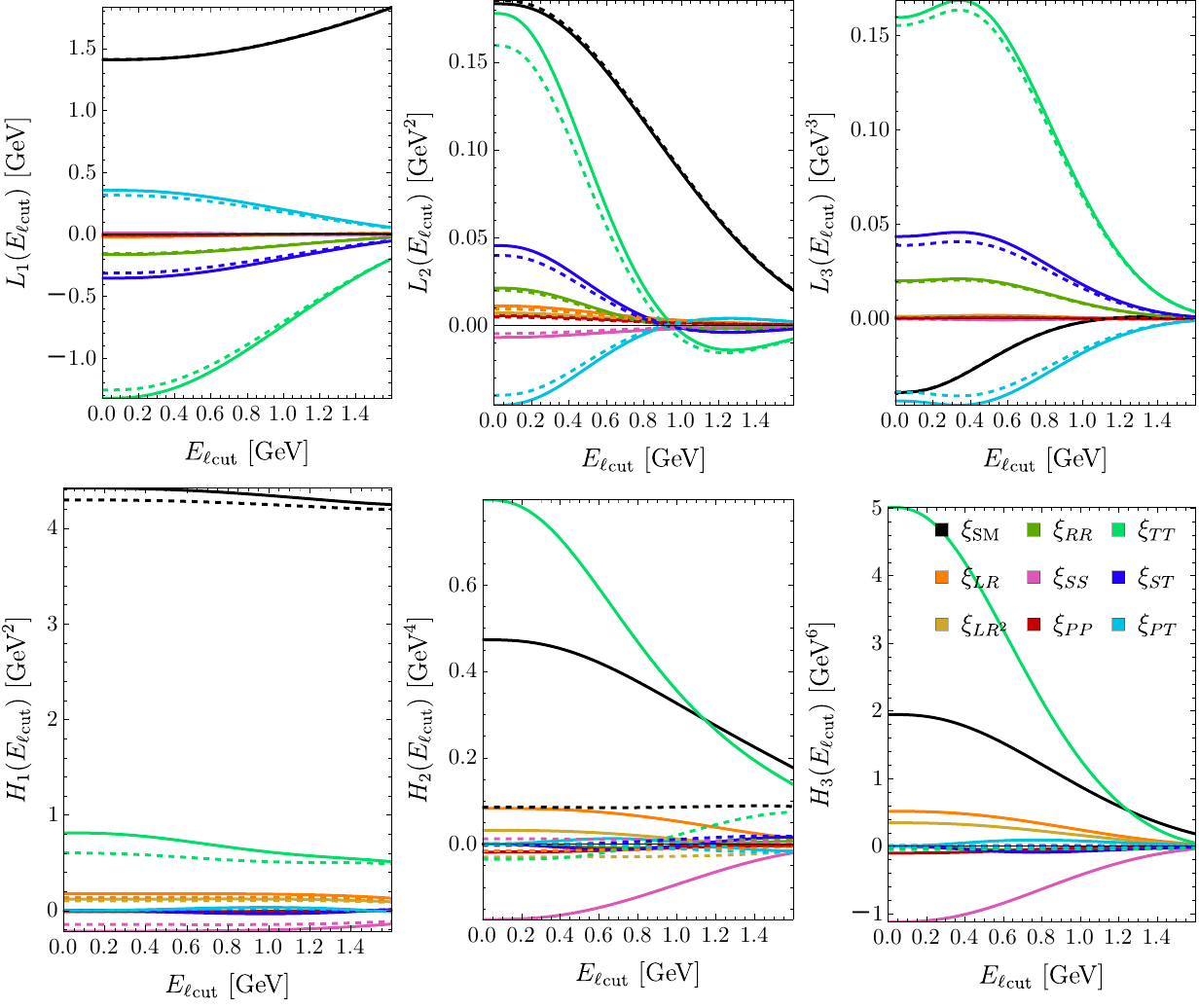}
    \caption{\small Plots for the several $\xi_i$ contributions to the lepton energy moments (upper panels) and hadronic invariant mass moments (lower panels). Different colours stand for different $\xi_i$, defined in~\eqref{eq:xidec}. Dashed lines stand for LO while solid lines for NLO.}
    \label{fig:plotLH}
\end{figure}

\begin{figure}
    \centering
    \includegraphics[width=\textwidth]{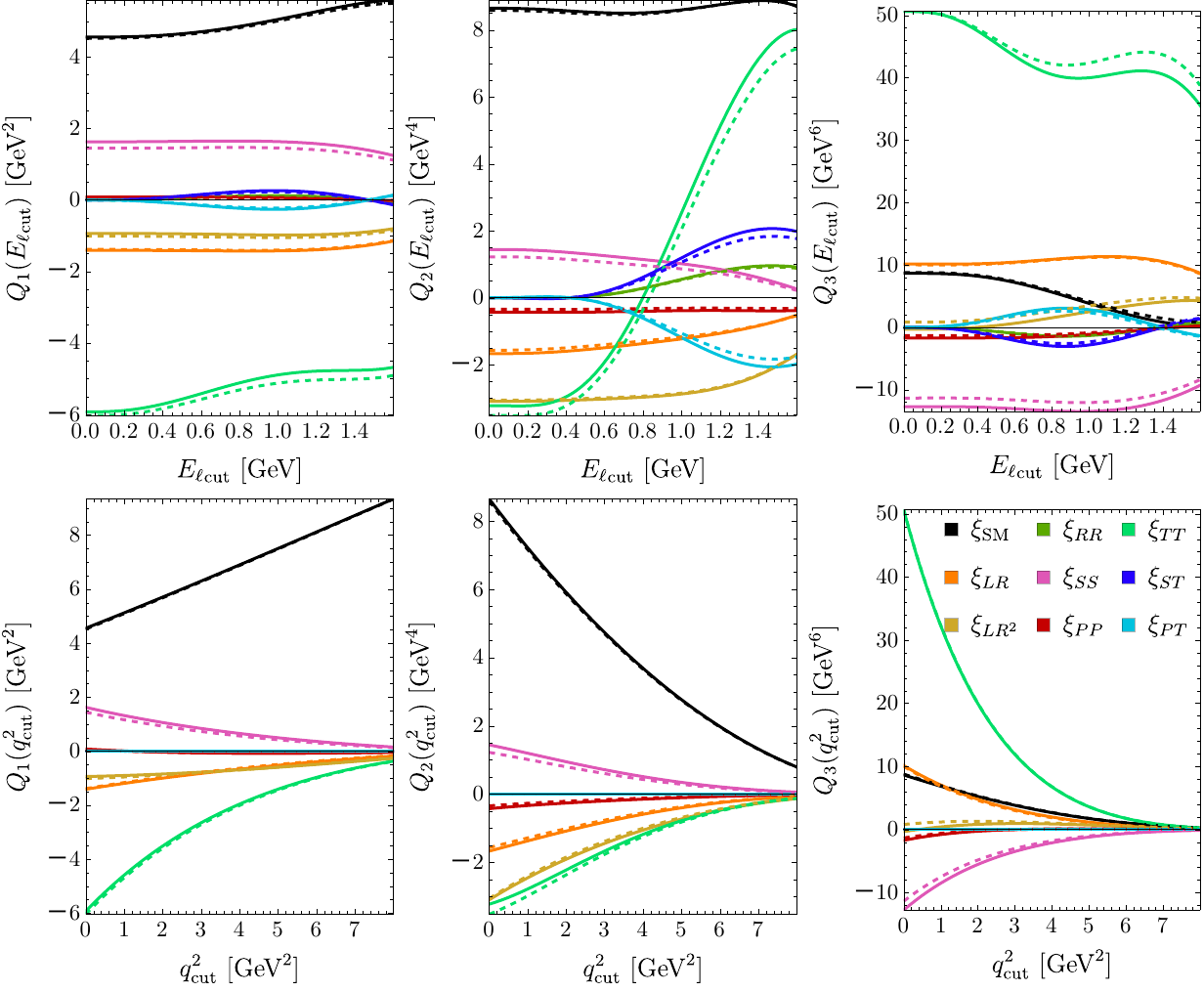}
    \caption{\small Plots for the several $\xi_i$ contributions to the $q^2$ moments with a lower cut on the lepton energy (upper panels) or on $q^2$ (lower panels). Different colours stand for different $\xi_i$, defined in~\eqref{eq:xidec}. Dashed lines stand for LO while solid lines for NLO.}
    \label{fig:plotQEcQ}
\end{figure}

In this section we provide a short numerical analysis of our results, to have a grasp on the impact of the $\mathcal{O}(\as)$ corrections with respect to the leading order terms.
In particular we show in Figures~\ref{fig:plotLH} and~\ref{fig:plotQEcQ} the four types of central moments defined in Section~\ref{sec:moms} as functions of their respective kinematical cut (${E_\ell}_{\rm cut}$ or $q^2_{\rm cut}$).
We used the following numerical inputs\footnote{The on-shell charm mass has been obtained converting the $\overline{\text{MS}}$ mass $m_c^{\overline{\text{MS}}} = 1.090~\text{GeV}$~\cite{Carvunis:2025vab} at the two-loop level.}
\begin{align}
\label{eq:inputs}
    m_b &= 4.8~\text{GeV}\,, \qquad m_c = 1.42~\text{GeV}\,,\nonumber\\
    \alpha_s^{(4)}(m_b) &= 0.215\,, \quad\qquad m_B = 5.279~\text{GeV}\,.
\end{align}
The contributions are split according to~\eqref{eq:xidec}. 
One can notice how the inclusion of $\mathcal{O}(\as)$ corrections is crucial in hadronic mass moments.
Furthermore it is interesting to see that the various coefficient functions in $q^2$ moments have drastically different behaviours if the cut is applied on $q^2$ or on $E_\ell$.
This reinforces the need for new measurements of these observables which could improve significantly the result of the global fit to inclusive observables~\cite{Carvunis:2025vab}.
\begin{figure}
    \centering
    \includegraphics[width=0.5\textwidth]{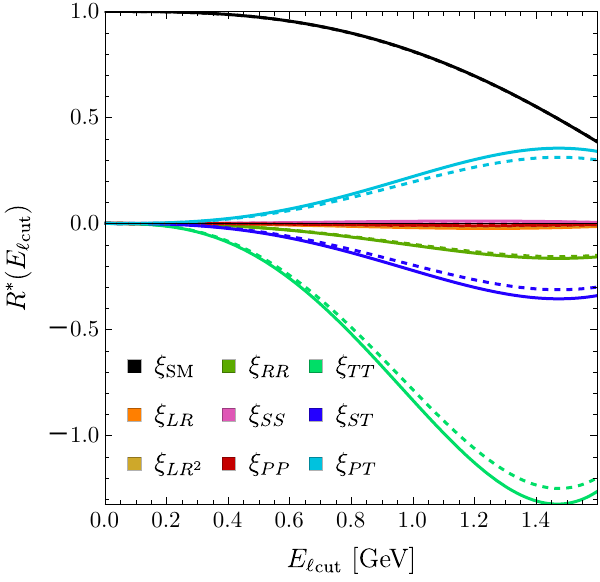}
    \caption{\small Plot for the several $\xi_i$ contributions to $R^*({E_\ell}_{\rm cut})$. Different colours stand for different $\xi_i$, defined in~\eqref{eq:xidec}. Dashed lines stand for LO while solid lines for NLO.}
    \label{fig:plotRstar}
\end{figure}
In Figure~\ref{fig:plotRstar} we show $R^*({E_\ell}_{\rm cut})$ where the $\as$ corrections are well under control for all the NP components.
We compared our results with the few analytic expressions in appendix C of Ref.~\cite{Aquila:2005hq}, corresponding to some of the building blocks for (not normalized) hadronic invariant mass moments in the SM.
We fully agree with those expressions.
Concerning NP, the comparison with the numerical tables of Ref.~\cite{Fael:2022wfc} is possible only after converting our results to the kinetic scheme.
This is done in appendix~\ref{sec:app} where we show numerical tables for the observables with the added $\mathcal{O}(\as)$ terms computed in this work.

\subsection{CLEO Data on $q^2$ Moments with Lepton Energy Cut}
\label{sec:CLEO}
In this section we briefly show the impact of the $\mathcal{O}(\as)$ corrections to $Q_{1,2}({E_\ell}_{\rm cut})$ in the Standard Model.
The goal is to update the comparison between the SM prediction and the measurement performed by CLEO in 2004~\cite{CLEO:2004bqt}, as these observables showed puzzling tensions with the tree level prediction~\cite{Finauri:2025ost}.
We recall the experimental results\footnote{Statistic and systematic uncertainties are summed in quadrature.} from Ref.~\cite{CLEO:2004bqt}
\begin{align}
\label{eq:CLEOdata}
\langle q^2 \rangle(E_\ell > 1~\text{GeV}) &= 4.89 \pm 0.14~\text{GeV}^2\,, \nonumber\\
\langle q^2 \rangle(E_\ell > 1.5~\text{GeV}) &= 5.29 \pm 0.12~\text{GeV}^2\,, \nonumber\\
\langle (q^2-\langle q^2\rangle)^2 \rangle(E_\ell > 1~\text{GeV}) &= 2.852 \pm 0.047~\text{GeV}^4\,, \nonumber\\
\langle (q^2-\langle q^2\rangle)^2 \rangle(E_\ell > 1.5~\text{GeV}) &= 2.879 \pm 0.050~\text{GeV}^4\,.
\end{align}
Using the quark masses values~\eqref{eq:inputs} we get the following HQE expansion of the observables
\begin{align}
\label{eq:thpredq2Ecut}
    Q_1(1~\text{GeV}) &= (5.01+0.20\as)~\text{GeV}^2 -0.013 \mu^2_\pi -0.56 \mu^2_G +0.11 \frac{\rho^3_{LS}}{\text{GeV}} -0.88 \frac{\rho^3_D}{\text{GeV}}\,,\nonumber\\
    Q_1(1.5~\text{GeV}) &= (5.47+0.36\as)~\text{GeV}^2 -0.096 \mu^2_\pi -0.78 \mu^2_G +0.096 \frac{\rho^3_{LS}}{\text{GeV}} -1.23 \frac{\rho^3_D}{\text{GeV}}\,,\nonumber\\
    Q_2(1~\text{GeV}) &= \left[ (8.59+0.077\as)~\text{GeV}^2 + 0.016 \mu^2_\pi -1.97\mu^2_G + 0.42 \frac{\rho^3_{LS}}{\text{GeV}}-4.85 \frac{\rho^3_D}{\text{GeV}} \right]\text{GeV}^2 \,,\nonumber\\
    Q_2(1.5~\text{GeV}) &= \left[ (8.86-0.041\as)~\text{GeV}^2 - 0.14 \mu^2_\pi +2.11\mu^2_G + 0.31 \frac{\rho^3_{LS}}{\text{GeV}} -6.06 \frac{\rho^3_D}{\text{GeV}}\,\right] \text{GeV}^2,
\end{align}
where we included power corrections for clarity~\cite{Finauri:2025ost}.
It is evident from~\eqref{eq:thpredq2Ecut} that the perturbative corrections computed in this work play no role in explaining the difference between the data~\eqref{eq:CLEOdata} and the SM prediction for $Q_2({E_\ell}_{\rm cut})$ pointed out in~\cite{Finauri:2025ost}.

\section{Summary}
\label{sec:conclusion}
In this work we presented for the first time analytic results for the first three kinematic moments of the inclusive decay $\bar{B} \to X_c \ell \nub$ at order $\mathcal{O}(\as)$ within the full basis of dimension-six WET operators. We find agreement with the partial results available in the literature~\cite{Aquila:2005hq,Alberti:2015qmj,Dassinger:2007pj}.
This result is particularly relevant as it was shown that the kinematic moments of inclusive $\bar{B} \to X_c \ell \nub$ can provide competitive bounds on NP \cite{Carvunis:2025vab}.
In this context a readily available closed analytical expressions for $\mathcal{O}(\alpha_s)$ corrections should be very useful, as it facilitates the phenomenology of inclusive $b \to c \ell \nub$ decays.
Furthermore we have confirmed the existing tension in the second central $q^2$ moment with a lower cut in the lepton energy \cite{Finauri:2025ost}, reinforcing the usefulness of an experimental update for this observable.

\subsubsection*{Acknowledgements}
We would like to thank Paolo Gambino for extensive discussions, and for the careful reading of the manuscript.
The work of GF is supported by the Italian Ministry of University and Research (MUR) and the European Union (EU) — Next Generation EU, Mission 4, Component 1, PRIN 2022, grant 2022N4W8WR, CUP D53D23002830006. 
The work of AC is supported by the \textit{Bundesministerium f\"ur Forschung, Technologie und Raumfahrt } -- BMFTR.

\appendix
\section{Numerical Tables}
\label{sec:app}
In this appendix we report the numerical values of the HQE expansion for the observables introduced in Section~\ref{sec:moms} separated into the various NP contributions, with all available perturbative and non-perturbative corrections~\cite{Carvunis:2025vab}.
Here we have converted the theoretical expressions from the on-shell scheme to the kinetic scheme~\cite{Bigi:1996si,Fael:2020iea} with cutoff scale $\mu_k = 1~\text{GeV}$ (except for the charm mass, which is converted to the $\overline{\text{MS}}$ scheme at a scale of 2 GeV).
We employ the same numerical inputs as Ref.~\cite{Carvunis:2025vab} such that the only difference are the $\mathcal{O}(\as)$ terms in $\xi_{SS}$, $\xi_{PP}$, $\xi_{TT}$, $\xi_{ST}$ and $\xi_{PT}$ (shown in blue).
When comparing our results with the ones of Refs.~\cite{Fael:2022wfc,Fael:2024fkt} we found the same pattern of discrepancies exhibited by the power corrections, as noted in Section 2.2 of~\cite{Carvunis:2025vab}.
\begin{table}[H]
    \centering
    \begin{adjustbox}{max width=\textwidth}
    \begin{tabular}{|c|rll|rll|rll|}
    \hline
 & \multicolumn{3}{c|}{$L_1 ~[10^{-2} \ \textrm{GeV}]$} & \multicolumn{3}{c|}{$L_2 ~ [10^{-2} \ \textrm{GeV}^2]$} & \multicolumn{3}{c|}{$L_3~[10^{-3} \ \textrm{GeV}^3]$} \\ \hline
\multirow{2}{*}{$~\xi_{SM}~$} & $157.276$&$-1.676_{\textrm{pow}}$&$-0.312_{\alpha_s} $ & $8.726$&$+0.279_{\textrm{pow}}$&$-0.034_{\alpha_s} $ & $-3.096$&$+3.201_{\textrm{pow}}$&$+0.986_{\alpha_s} $ \\ 
&& $-0.463_{\frac{\alpha_s}{m_b^2}} $ & $+0.084_{\alpha_s^2}$ & & $-0.192_{\frac{\alpha_s}{m_b^2}} $ & $+0.057_{\alpha_s^2}$ & & $-0.364_{\frac{\alpha_s}{m_b^2}} $ & $+0.672_{\alpha_s^2}$ \\ \hline
$~\xi_{RR}~$ & $-9.760$&$+1.114_{\textrm{pow}}$&$+0.208_{\alpha_s}$ & $0.004$&$-0.287_{\textrm{pow}}$&$+0.021_{\alpha_s}$ & $9.004$&$-1.541_{\textrm{pow}}$&$-0.431_{\alpha_s}$ \\ \hline
$~\xi_{LR}~$ & $-0.368$&$+0.606_{\textrm{pow}}$&$-0.197_{\alpha_s}$ & $0.279$&$+0.105_{\textrm{pow}}$&$-0.013_{\alpha_s}$ & $0.576$&$+0.235_{\textrm{pow}}$&$+0.047_{\alpha_s}$ \\ \hline
$~\xi_{LR^2}~$ & $-0.247$&$+0.427_{\textrm{pow}}$&$-0.128_{\alpha_s}$ & $0.186$&$+0.060_{\textrm{pow}}$&$-0.013_{\alpha_s}$ & $0.418$&$+0.087_{\textrm{pow}}$&$+0.040_{\alpha_s}$ \\ \hline
$~\xi_{TT}~$ & $-78.076$&$+9.286_{\textrm{pow}}$& \red{$+1.044_{\alpha_s}$} & $0.033$&$-1.071_{\textrm{pow}}$& \red{$+0.163_{\alpha_s}$} & $72.030$&$-6.863_{\textrm{pow}}$& \red{$-3.405_{\alpha_s}$} \\ \hline
$~\xi_{SS}~$ & $0.184$&$+3.006_{\textrm{pow}}$& \red{$+0.096_{\alpha_s}$} & $-0.139$&$+0.471_{\textrm{pow}}$& \red{$-0.015_{\alpha_s}$} & $-0.288$&$-2.420_{\textrm{pow}}$& \red{$+0.026_{\alpha_s}$}\\ \hline
$~\xi_{PP}~$ & $-0.184$&$+0.191_{\textrm{pow}}$& \red{$+0.288_{\alpha_s}$} & $0.139$&$+0.030_{\textrm{pow}}$& \red{$+0.051_{\alpha_s}$} & $0.288$&$-0.169_{\textrm{pow}}$& \red{$-0.172_{\alpha_s}$}\\ \hline
$~\xi_{ST}~$ & $-19.519$&$-2.147_{\textrm{pow}}$& \red{$-1.332_{\alpha_s}$} & $0.008$&$-1.988_{\textrm{pow}}$& \red{$+0.039_{\alpha_s}$} & $18.007$&$-3.218_{\textrm{pow}}$& \red{$+0.767_{\alpha_s}$}\\ \hline
$~\xi_{PT}~$ & $19.519$&$-0.025_{\textrm{pow}}$& \red{$+1.332_{\alpha_s}$} & $-0.008$&$+0.776_{\textrm{pow}}$& \red{$-0.039_{\alpha_s}$} & $-18.007$&$+0.502_{\textrm{pow}}$& \red{$-0.767_{\alpha_s}$}\\ \hline
\end{tabular}
    \end{adjustbox}
    \vspace{-0.3cm}
    \caption{\small Lepton energy moments with ${E_\ell}_{\rm cut} = 1~\textrm{GeV}$.}
    \label{tab:El_moments}
\end{table}

\begin{table}[H]
    \centering
    \begin{adjustbox}{max width=\textwidth}
    \begin{tabular}{|c|rll|rll|rll|}
    \hline
 & \multicolumn{3}{c|}{$H_1 ~ [10^{-1} \ \textrm{GeV}^2]$} & \multicolumn{3}{c|}{$H_2 ~[10^{-1} \ \textrm{GeV}^4]$} & \multicolumn{3}{c|}{$H_3~ [ 10^{-1} \ \textrm{GeV}^6]$} \\ \hline
\multirow{2}{*}{$~\xi_{SM}~$} & $42.959$&$+0.031_{\textrm{pow}}$&$+0.491_{\alpha_s} $ & $2.236$&$+7.089_{\textrm{pow}}$&$+2.043_{\alpha_s} $ & $-0.213$&$+47.209_{\textrm{pow}}$&$-3.632_{\alpha_s} $ \\ 
&& $+0.418_{\frac{\alpha_s}{m_b^2}} $ & $-0.113_{\alpha_s^2}$ & & $-1.146_{\frac{\alpha_s}{m_b^2}} $ & $+1.066_{\alpha_s^2}$ & & $-4.032_{\frac{\alpha_s}{m_b^2}} $ & $-3.975_{\alpha_s^2}$ \\ \hline
$~\xi_{RR}~$ & $-0.213$&$-0.085_{\textrm{pow}}$&$-0.052_{\alpha_s}$ & $0.146$&$-1.072_{\textrm{pow}}$&$-0.054_{\alpha_s}$ & $0.059$&$-0.499_{\textrm{pow}}$&$-0.491_{\alpha_s}$ \\ \hline
$~\xi_{LR}~$ & $2.135$&$-2.023_{\textrm{pow}}$&$+0.019_{\alpha_s}$ & $-0.365$&$+5.087_{\textrm{pow}}$&$+0.011_{\alpha_s}$ & $-0.405$&$+8.232_{\textrm{pow}}$&$-0.395_{\alpha_s}$ \\ \hline
$~\xi_{LR^2}~$ & $1.434$&$-1.477_{\textrm{pow}}$&$-0.011_{\alpha_s}$ & $-0.701$&$+4.301_{\textrm{pow}}$&$+0.003_{\alpha_s}$ & $-0.038$&$+2.071_{\textrm{pow}}$&$-0.266_{\alpha_s}$ \\ \hline
$~\xi_{TT}~$ & $7.917$&$+6.888_{\textrm{pow}}$& \red{$+0.134_{\alpha_s}$}& $0.492$&$+12.225_{\textrm{pow}}$& \red{$+0.697_{\alpha_s}$} & $-1.625$&$+37.872_{\textrm{pow}}$& \red{$+1.442_{\alpha_s}$}\\ \hline
$~\xi_{SS}~$ & $-2.271$&$-2.994_{\textrm{pow}}$& \red{$-0.501_{\alpha_s}$} & $0.267$&$-2.385_{\textrm{pow}}$& \red{$-0.255_{\alpha_s}$} & $0.464$&$-9.426_{\textrm{pow}}$& \red{$-1.363_{\alpha_s}$}\\ \hline
$~\xi_{PP}~$ & $-0.136$&$+0.021_{\textrm{pow}}$& \red{$-0.339_{\alpha_s}$} & $-0.098$&$-0.793_{\textrm{pow}}$& \red{$-0.732_{\alpha_s}$} & $0.059$&$-1.135_{\textrm{pow}}$& \red{$-2.430_{\alpha_s}$} \\ \hline
$~\xi_{ST}~$ & $-0.427$&$+0.542_{\textrm{pow}}$& \red{$-0.123_{\alpha_s}$} & $0.292$&$-2.504_{\textrm{pow}}$& \red{$-0.003_{\alpha_s}$} & $0.117$&$-1.537_{\textrm{pow}}$& \red{$-0.535_{\alpha_s}$} \\ \hline
$~\xi_{PT}~$ & $0.427$&$-0.258_{\textrm{pow}}$& \red{$+0.123_{\alpha_s}$} & $-0.292$&$+1.367_{\textrm{pow}}$& \red{$+0.003_{\alpha_s}$} & $-0.117$&$+0.863_{\textrm{pow}}$& \red{$+0.535_{\alpha_s}$}\\ \hline
    \end{tabular}
    \end{adjustbox}
    \vspace{-0.3cm}
    \caption{\small Hadronic mass moments with ${E_\ell}_{\rm cut}= 1~\textrm{GeV}$.}
    \label{tab:MX_moments}
\end{table}

\begin{table}[H]
    \centering
    \begin{adjustbox}{max width=\textwidth}
    \begin{tabular}{|c|rll|rll|rll|}
    \hline
 & \multicolumn{3}{c|}{$Q_1 ~[\textrm{GeV}^2]$} & \multicolumn{3}{c|}{$Q_2 ~[\textrm{GeV}^4]$} & \multicolumn{3}{c|}{$Q_3~[\textrm{GeV}^6]$} \\ \hline
\multirow{2}{*}{$~\xi_{SM}~$} & $7.077$&$-0.425_{\textrm{pow}}$&$+0.012_{\alpha_s} $ & $4.293$&$-1.639_{\textrm{pow}}$&$+0.060_{\alpha_s} $ & $3.795$&$-4.475_{\textrm{pow}}$&$+0.477_{\alpha_s} $ \\ 
&$-0.028_{\frac{\alpha_s}{m_b^2}} $ & $-0.024_{\frac{\alpha_s}{m_b^3}} $ & $+0.041_{\alpha_s^2}$ & $-0.050_{\frac{\alpha_s}{m_b^2}} $ & $-0.041_{\frac{\alpha_s}{m_b^3}} $ & $+0.168_{\alpha_s^2}$ & $+0.040_{\frac{\alpha_s}{m_b^2}} $ & $+0.177_{\frac{\alpha_s}{m_b^3}} $ & $+0.578_{\alpha_s^2}$ \\ \hline
$~\xi_{LR}~$ & $-0.693$&$+0.101_{\textrm{pow}}$&$+0.006_{\alpha_s}$ & $-0.769$&$+0.233_{\textrm{pow}}$&$+0.092_{\alpha_s}$ & $2.117$&$-0.768_{\textrm{pow}}$&$+0.165_{\alpha_s}$ \\ \hline
$~\xi_{LR^2}~$ & $-0.681$&$+0.129_{\textrm{pow}}$&$+0.008_{\alpha_s}$ & $-1.236$&$+0.402_{\textrm{pow}}$&$+0.101_{\alpha_s}$ & $0.481$&$-0.129_{\textrm{pow}}$&$+0.360_{\alpha_s}$ \\ \hline
$~\xi_{TT}~$ & $-2.179$&$+0.484_{\textrm{pow}}$& \red{$+0.036_{\alpha_s}$} & $-1.598$&$+1.228_{\textrm{pow}}$& \red{$-0.019_{\alpha_s}$} & $8.239$&$-1.608_{\textrm{pow}}$& \red{$-0.696_{\alpha_s}$}\\ \hline
$~\xi_{SS}~$ & $0.619$&$+0.677_{\textrm{pow}}$& \red{$+0.104_{\alpha_s}$} & $0.584$&$+2.189_{\textrm{pow}}$& \red{$+0.091_{\alpha_s}$} & $-2.088$&$+4.671_{\textrm{pow}}$& \red{$-0.280_{\alpha_s}$}\\ \hline
$~\xi_{PP}~$ & $-0.074$&$+0.004_{\textrm{pow}}$& \red{$+0.000_{\alpha_s}$} & $-0.185$&$+0.029_{\textrm{pow}}$& \red{$+0.058_{\alpha_s}$} & $0.028$&$-0.018_{\textrm{pow}}$& \red{$+0.162_{\alpha_s}$}\\ \hline
    \end{tabular}
    \end{adjustbox}
    \vspace{-0.3cm}
    \caption{\small $q^2$ moments with $q^2_{\rm cut}= 4~\textrm{GeV}^2$. We have omitted the lines for $\xi_{RR}=\xi_{ST}=\xi_{PT}=0$.}
    \label{tab:q2_moments}
\end{table}
\begin{table}[H]
    \centering
    \begin{tabular}{|l|rlll|}
    \hline 
    &\multicolumn{4}{c|}{$10 \times R^{*}$} \\ \hline
$~\xi_{SM}~$ & $8.175$ & $-0.125_{\textrm{pow}}$ & $-0.001_{\alpha_s} $ & $-0.016_{\alpha_s/m_b^2} -0.004_{\alpha_s^2}$ \\ \hline 
$~\xi_{RR}~$ & $-1.039$&$+0.020_{\textrm{pow}}$&$+0.003_{\alpha_s}$ & \\ \hline
$~\xi_{LR}~$ & $-0.174$&$+0.038_{\textrm{pow}}$&$-0.033_{\alpha_s}$ & \\ \hline
$~\xi_{LR^2}~$ & $-0.113$&$+0.033_{\textrm{pow}}$&$-0.019_{\alpha_s}$ & \\ \hline
$~\xi_{TT}~$ & $-8.316$&$+0.024_{\textrm{pow}}$& \red{$-0.149_{\alpha_s}$}& \\ \hline
$~\xi_{SS}~$ & $0.087$&$+0.150_{\textrm{pow}}$ & \red{$+0.038_{\alpha_s}$}& \\ \hline
$~\xi_{PP}~$ & $-0.087$&$-0.002_{\textrm{pow}}$& \red{$+0.005_{\alpha_s}$}& \\ \hline
$~\xi_{ST}~$ & $-2.079$&$-0.081_{\textrm{pow}}$& \red{$-0.172_{\alpha_s}$}& \\ \hline
$~\xi_{PT}~$ & $2.079$&$+0.063_{\textrm{pow}}$& \red{$+0.172_{\alpha_s}$}& \\ \hline
    \end{tabular}
    \caption{\small $R^*$ with ${E_\ell}_{\rm cut} = 1~\text{GeV}$.}
    \label{tab:Rstar}
\end{table}
The inclusive semileptonic decay rate is parametrized as
\begin{equation}
    \Gamma = \Gamma_0 \biggl(\zeta_{\rm SM} + a_R \cos(\delta_R) \zeta_{LR} + a_R^2 \zeta_{RR} + a_S^2 \zeta_{SS} + a_P^2 \zeta_{PP} + a_T^2 \zeta_{TT} \biggr)\,,
\end{equation}
where $\Gamma_0$ is defined in~\eqref{eq:Gamma0}.
\begin{table}[H]
    \centering
    \begin{adjustbox}{max width=\textwidth}
    \begin{tabular}{|c|rlll|}
    \hline 
    &\multicolumn{4}{c|}{$10 \times \Gamma/\Gamma_0$} \\ \hline
$~\zeta_{SM}~$ & $6.580$ & $-0.461_{\textrm{pow}}$ & $-0.560_{\alpha_s} $ & $-0.009_{\alpha_s/m_b^2} -0.034_{\alpha_s/m_b^3} -0.105_{\alpha_s^2}+0.003_{\alpha_s^3}+0.058_{\alpha_{\rm em}}$ \\ \hline 
$~\zeta_{RR}~$ & $6.580$&$-0.461_{\textrm{pow}}$&$-0.560_{\alpha_s}$ & \\ \hline
$~\zeta_{LR}~$ & $-4.280$&$+0.634_{\textrm{pow}}$&$+0.464_{\alpha_s}$ & \\ \hline
$~\zeta_{TT}~$ & $78.964$&$-7.865_{\textrm{pow}}$& \red{$-5.694_{\alpha_s}$}& \\ \hline
$~\zeta_{SS}~$ & $5.430$&$+0.240_{\textrm{pow}}$& \red{$+0.354_{\alpha_s}$}& \\ \hline
$~\zeta_{PP}~$ & $1.150$&$-0.118_{\textrm{pow}}$& \red{$-0.276_{\alpha_s}$}& \\ \hline
    \end{tabular}
    \end{adjustbox}
    \vspace{-0.3cm}
    \caption{\small Inclusive semileptonic decay width $\Gamma$.}
    \label{tab:Gammatot}
\end{table}

\bibliographystyle{JHEP} % bst file
\bibliography{refs.bib}

\end{document}